\documentclass[useAMS,usenatbib]{mn2e}

\title[HOD of Black Holes: Dependence on Mass]{The Halo Occupation Distribution of
Black Holes: Dependence on Mass}

\author[Colin Degraf et al.]  {Colin Degraf$^{1}$,
       Matthew Oborski$^{1}$, Tiziana Di Matteo$^{1}$, Suchetana
       Chatterjee$^2$, 
       \newauthor
       Daisuke Nagai$^{2,3}$, Zheng Zheng$^3$, Jonathan Richardson$^2$ \\ 
       $^{1}$ McWilliams Center for
       Cosmology, Carnegie Mellon University, 5000 Forbes Avenue, Pittsburgh,
       PA 15213, USA\\
       $^2$ Department of Astronomy, Yale University, New Haven, CT 06520 USA\\
       $^3$ Department of Physics, Yale University, New Haven, CT 06520 USA\\
}

\usepackage{ graphicx}
\usepackage{ subfigure}
\usepackage{multirow}

\begin{document}

\date{Accepted 20?? ???? ??.
      Received 20?? ???? ??;
      in original form 20??  xx}
\pagerange{\pageref{firstpage}--\pageref{lastpage}}
\pubyear{20??}

\maketitle

\begin{abstract}
  We investigate the halo occupation distribution (HOD) of black holes within
  a hydrodynamic cosmological simulation that directly follows black hole
  growth. Similar to the HOD of galaxies/subhalos, we find that the black
  hole occupation number can be described by the form $N_{\rm{BH}} \propto 1+
  ({M_{\rm{Host}}})^{\alpha}$ where $\alpha$ evolves mildly with redshift
  indicating that a given mass halo ($M_{\rm{Host}}$) at low redshift tends to host
  fewer BHs than at high redshift (as expected as a result of galaxy and BH
  mergers).  We further divide the occupation number into contributions from
  black holes residing in central and satellite galaxies within a halo. The distribution of
  $M_{\rm{BH}}$ within halos tends to consist of a single massive BH (distributed
  about a peak mass strongly correlated with $M_{\rm{Host}}$), and a
  collection of relatively low-mass secondary BHs, with weaker correlation
  with $M_{\rm{Host}}$.  We also examine the spatial distribution of BHs
  within their host halos, and find they typically follow a power-law radial
  distribution (i.e. much more centrally concentrated than the subhalo
  distribution).  Finally, we characterize the host mass for which BH growth
  is feedback dominated (e.g. star formation quenched). We show that
  halos with $M_{\rm{Host}} > 3 \times 10^{12} M_\odot$ have primary BHs that
  are feedback dominated by $z \sim 3$ with lower mass halos becoming
  increasingly more affected at lower redshift.
\end{abstract}

\section{Introduction}

Supermassive black holes have been found to be at the center of most galaxies
 \citep{1995ARA&A..33..581K}, and the correlation between these central black
holes and their host galaxy properties have been extensively studied 
\citep[see, e.g.][]{1998AJ....115.2285M,
  2000ApJ...539L...9F, 2000ApJ...539L..13G, Tremaine2002,
  2007ApJ...655...77G}.  Gas accretion onto supermassive black holes leads to
 AGN and quasar activity, producing bright objects idea for observations.  One of the primary means of observationally studying
 quasars is by looking at their clustering
properties, including redshift evolution \citep[e.g.][]{LaFranca1998, Porciani2004, Croom2005, Shen2007, Myers2007,
  daAngela2008, Shen2009, Ross2009} and luminosity dependence \citep[see,
e.g.,][]{Croom2005, Myers2007, daAngela2008, Shen2009}.  By looking at the
clustering strength of different quasar populations, one can estimate the
mass of the typical host halo \citep{Lidz2006,Ross2009,Bonoli2009,
  Shen2009}, thereby getting
a sense of how black holes populate halos.  This information can then be used
to infer details about
several properties, such as quasar lifetimes \citep{HaimanHui2001, MartiniWeinberg2001},
which can be constrained by comparing the observed quasar number density to the predicted
density of the halo mass-function prediction using the typical host halos.   

Recently, quasar clustering studies have found evidence for a bias in the
small-scale correlation function \citep{Hennawi2006, Myers2007II, Myers2008},
and \citet{DeGraf2010Clustering} used cosmological hydrodynamic simulations to
investigate BH clustering, finding that the existence of multiple BHs within
individual galaxies has a significant effect on the small-scale correlation
function which could explain this observed small-scale excess.  However, this was an indirect means of exploring
the relation between BHs and their typical host halos, and a more direct
investigation is necessary to fully understand how BHs populate halos.

With the aid of simulations, the opposite approach can be taken for such a
direct investigation: instead of using clustering to predict the halos occupied by BHs, the DM halos
can be directly probed to get BH occupation properties, which can then be extended to
explore clustering properties.  This technique has been used
extensively for galaxies in the form of the Halo Occupation Distribution (HOD) \citep[see,
  e.g.][]{BerlindWeinberg2002, Kravtsov2004, Zheng2005}.  At the most basic
level, an HOD model
characterizes the number of objects within halos as a function of the halo
mass, and how they are
spatially distributed within the halo.  Given these simple statistical
distributions, the HOD model can be used to populate halos in N-body
simulations \citep[e.g.][]{Benson2000, Berlind2003, Brown2008} and, assuming the occupation
distribution is independent of the large scale environment \citep{LemsonKauffmann1999,Berlind2003}, can be used to analytically calculate
the clustering statistics for a given cosmological model \citep[see,
  e.g.][]{Seljak2000, BerlindWeinberg2002}.
The galaxy HOD model can then be extended further by looking at how the occupation
properties depend upon various galaxy parameters, such as galaxy luminosity,
color, or morphology, which can be used to better understand the physics of
galaxy formation and evolution \citep[e.g.][]{Yoshikawa2001, Berlind2003,
  Zehavi2005, Zheng2007,
  ReidSpergel2009, Zehavi2010}.

Despite its overall success for galaxies, the HOD technique
has not been applied to black holes.  In this paper we extend the work done
in \citet{DeGraf2010Clustering} by directly investigating how BHs populate
dark matter halos using the HOD formalism.  In addition to characterizing the
occupation number of BHs in DM halos, we investigate the distribution of BH
masses within the halo, as well as their spatial distribution among the component
subhalos.  Additionally, in an upcoming paper (Chatterjee et al., in prep) we will further extend this
model to incorporate the luminosity dependencies of the black holes.  By providing these details of a BH HOD model from a
hydrodynamic simulation, we hope to improve the techniques available for
both semi-analytic BH models, and theoretical studies of BH clustering. 

In Section \ref{sec:Method} we describe the simulation used, with particular emphasis on how the black holes are modeled.  In Section \ref{sec:Results} 
we investigate the BH occupation number both for halos and for central and
satellite galaxies (\ref{sec:occupation}), the distribution of BH masses as a function of host
halo mass (\ref{sec:cmf}), and the spatial distribution of the BHs within their parent
halos (\ref{sec:radialdist}).  Finally, in section (\ref{sec:feedback}) we
look at when the feedback from the BH begins to suppress further BH growth, and
we summarize our results in Section 4.

\section{Method}
\label{sec:Method}

\subsection{Numerical simulation}

In this study, we analyse the set of simulations published in
\citet{DiMatteo2008}. Here we present a brief summary of the simulation
code and the method used. We refer the reader to \citet{DiMatteo2008}
for all details.

 The code we use is the massively parallel cosmological TreePM--SPH code
 {\small Gadget2} \citep{2005MNRAS.364.1105S}, with the addition of a multi--phase
 modeling of the ISM, which allows treatment of star formation
 \citep{2003MNRAS.339..289S}, and black hole accretion and associated feedback
 processes \citep{2005MNRAS.361..776S, 2005Natur...433..604D}. Detailed studies of the prescription for accretion and associated
  feedback from massive black holes and associated predictions have been presented in \citet{Sijacki2007, DiMatteo2008, Colberg2008, Croft2009,
    DeGraf2010}.  Important for our discussion is that the model has been
  shown to reproduce remarkably well both the observed $M_{\rm{BH}}-\sigma$
  relation and total black hole mass density $\rho_{\rm{BH}}$ \citep{DiMatteo2008}, as well as the quasar
  luminosity functions and its evolution in optical, soft and hard X-ray band \citep{DeGraf2010}. Thus the model, within its intrinsic limitations, appears to
  serve as a fair standard for representing growth, activity, and evolution
  of massive black holes in numerical simulations, at least within
  the context of cosmological growth of black holes and not the detailed accretion
  physics (the detailed treatment of which is completely infeasible in
  cosmological simulations). We note also that at least two independent teams
  \citep{Booth2009, Johansson2008} now have also adopted the same modeling for
  black hole accretion, feedback and BH mergers in the context of hydrodynamic
  simulations. These independent works, in particular the cosmological
  simulations by \citet{Booth2009} (part of the OWLS program) have 
  fully and independently explored the parameter space of the reference model
  of \citet{DiMatteo2008}, as well as variations to some prescriptions. This
  large body of already existing work and investigations make this particular
  model somewhat of a good choice for further study.

  Within the simulation, black holes are simulated with collisionless
  particles that are created in newly emerging and resolved groups/galaxies.
  To find these groups, a friends--of--friends group finder is called at
  regular intervals on the fly (in time intervals equally spaced in $\log_{10} (a)$,
  with $\Delta \log{a} = \log{1.25}$), finding groups based on particle
  separations below a specified cutoff.  Each group with a mass above $5
  \times 10^{10} h^{-1}$\,M$_\odot$ that does not already contain a black hole
  is provided with one by converting its densest particle into a sink particle
  with a seed mass of $ M_{\rm{BH,seed}} = 5 \times 10^5 h^{-1}$\,M$_\odot$. This
  seeding prescription was selected to reasonably match the expected formation
  of supermassive black holes either by collapse of a supermassive star to a
  BH with $M_{\rm{BH}} \sim M_{\rm{seed}}$ \citep[e.g.][]{BrommLoeb2003,
    Begelman2006} or by Pop III stars collapsing into $\sim 10^2 M_\odot$ BHs
  at $z \sim 30$ followed by exponential growth \citep{Bromm2004,
    Yoshida2006}, reaching $\sim M_{\rm{seed}}$ by the time the group reaches
  $\sim 10^{10} M_{\odot}$.  After insertion, the black hole particle grows in
  mass via both accretion of surrounding gas and by merging with other black
  holes.  The gas accretion is modeled according to $\dot{M}_{\rm BH} = \frac
  {4 \pi G^2 M_{\rm BH}^2 \rho}{(c_s^2 + v^2)^{3/2}}$
  \citep{1939PCPS...35..405H, 1944MNRAS.104..273B, 1952MNRAS.112..195B}, where
  $\rho$ is the local gas density, $c_s$ is the local sound speed, and $v$ is
  the velocity of the BH relative to the surrounding gas.  Note that to allow
  for the initial rapid BH growth necessary to produce supermassive BHs of
  $\sim 10^9 M_\odot$ at early time ($z \sim 6$), we allow for mildly
  super-Eddington accretion \citep[consistent with models of,
  e.g.,][]{VolonteriRees2006, Begelman2006}, but limit super-Eddington
  accreion to a maximum of $3 \times \dot{M}_{\rm{Edd}}$ to prevent
  artificially high values.

	The accretion rate of each black hole is used to compute the
	bolometric luminosity, $L = \eta \dot{M}_{\rm BH} c^2$
	\citep{1973A&A....24..337S}.  Here $\eta$ is the radiative efficiency,
	and it is fixed at 0.1 throughout the simulation and this analysis.
	Some coupling between the liberated luminosity and the surrounding gas
	is expected, modeled in the simulation by isotropically depositing the
	5 per cent of the luminosity as thermal energy to the local black hole
	kernal.  This parameter is fixed at 5 per cent based
	on earlier galaxy merger simulations such that the normalization
	of the $M_{\rm{BH}}-\sigma$ relation is reproduced \citep{2005Natur...433..604D}.

\begin{table}
\caption{Numerical Parameters}
\begin{tabular}{c c c c c}

  \hline
  \hline
  
  Boxsize & $N_p$ & $m_{\rm DM}$ & $m_{\rm gas}$ & $\epsilon$ \\
  $h^{-1} {\rm Mpc}$ & & $h^{-1} M_{\odot}$ & $h^{-1} M_{\odot}$ & $h^{-1}
   {\rm kpc}$ \\
  
  \hline
  
  33.75 & $2 \times 486^3$ & $2.75 \times 10^7$ & $4.24 \times 10^6$ & 2.73 \\
  
\hline

\multicolumn{5}{l}{$N_p$: Total number of particles} \\
\multicolumn{5}{l}{$m_{\rm DM}$: Mass of dark matter particles} \\
\multicolumn{5}{l}{$m_{\rm gas}$: Initial mass of gas particles} \\
\multicolumn{5}{l}{$\epsilon$: Comoving gravitational softening length} \\

\end{tabular}
\label{param}
\end{table}

The other means of black hole growth is via mergers.  When dark matter halos
merge into a single halo, the black holes typically fall toward the center of
the new halo, eventually merging with one another.  For these cosmological
volumes, it is not possible to directly calculate the details of the infalling
BHs at the smallest scales, so a sub-resolution merger prescription is used.
Since the merging BHs are typically found in a gaseous environment at the
center of a galaxy, we assume that the final coalescence will be rapid
\citep{MakinoFunato2004, Escala2004, Mayer2007}.  Thus our BHs merge when they
come within the spatial resolution of the simulation.  However, to prevent
merging of BHs which are rapidly passing one another, mergers are permitted
only if the velocity of the BHs relative to one another is small (comparable to the local sound
speed).

In addition to having been used in galaxy merger simulations to investigate
the regulation of BH growth and correlation with host galaxies
\citep{2005Natur...433..604D, Robertson2006, 2007ApJ...669...67H},
\citet{DiMatteo2008} previously investigated the validity of this method of
modeling black holes in these cosmological simulations, finding that in
addition to producing an $M_{\rm{BH}} - \sigma$ relation that matches
observations, the black hole mass density matches values inferred from the
integrated x-ray luminosity function \citep{Shankar2004, 2004MNRAS.351..169M}
and the accretion rate density is consistent with the constraints of
\citet{2007ApJ...654..731H}.

\subsection{Simulation parameters}
The simulation analysed in this paper populates $2 \times 486^3$ particles in
a moderate volume of side length $33.75 h^{-1} {\rm Mpc}$ (See Table
\ref{param} for additional simulation parameters). This moderate
boxsize prevents the simulation from being run below $z\sim 1$ to keep the
fundamental mode linear, but provides a large enough scale to produce
statistically significant quasar populations. The limitation on the boxsize
is necessary to allow for appropriate resolution to carry out the subgrid
physics in a converged regime (for further details on the simulation methods,
parameters and convergence studies see \citet{DiMatteo2008}).

\subsection{Subgroup finder algorithm}
In addition to the on-the-fly friends-of-friends algorithm used to identify
groups, a modified version of the SUBFIND algorithm \citep{Springel2001} was
run on the FoF-identified groups to determine the component subgroups
(\textit{i.e.} galaxies) within each group.  These subgroups are defined as
locally overdense, self-bound particle groups.  To identify these regions, the
algorithm analyzes each particle within the parent group in order of decreasing density.  For each particle
\textit{i}, the density of the 32 nearest neighbors are checked.  If
none are denser than particle \textit{i}, it forms the basis for a new
subgroup.  If a single particle denser than \textit{i} is found, or if the
closest two denser particles belong to the same subgroup, particle \textit{i}
is assumed to be a member of that subgroup.  If the two nearest particles
denser than \textit{i} are members of different subgroups, these two subgroups
are stored as subgroup candidates, and are then joined into a new subgroup
also containing \textit{i}.  After checking each particle in this manner,
particles are checked for gravitational binding within their parent subgroup based on their
position relative to the position of the most bound particle and the velocity
relative to the mean velocity of particles in the group.  Any particle with
positive total energy is considered unbound, and is removed from the subgroup,
leaving the group divided up into its component subgroups (galaxies).

\section{Results}
\label{sec:Results}

\subsection{Black Hole Occupation Number}
\label{sec:occupation}
\begin{figure*}
\centering
\includegraphics[width=16cm]{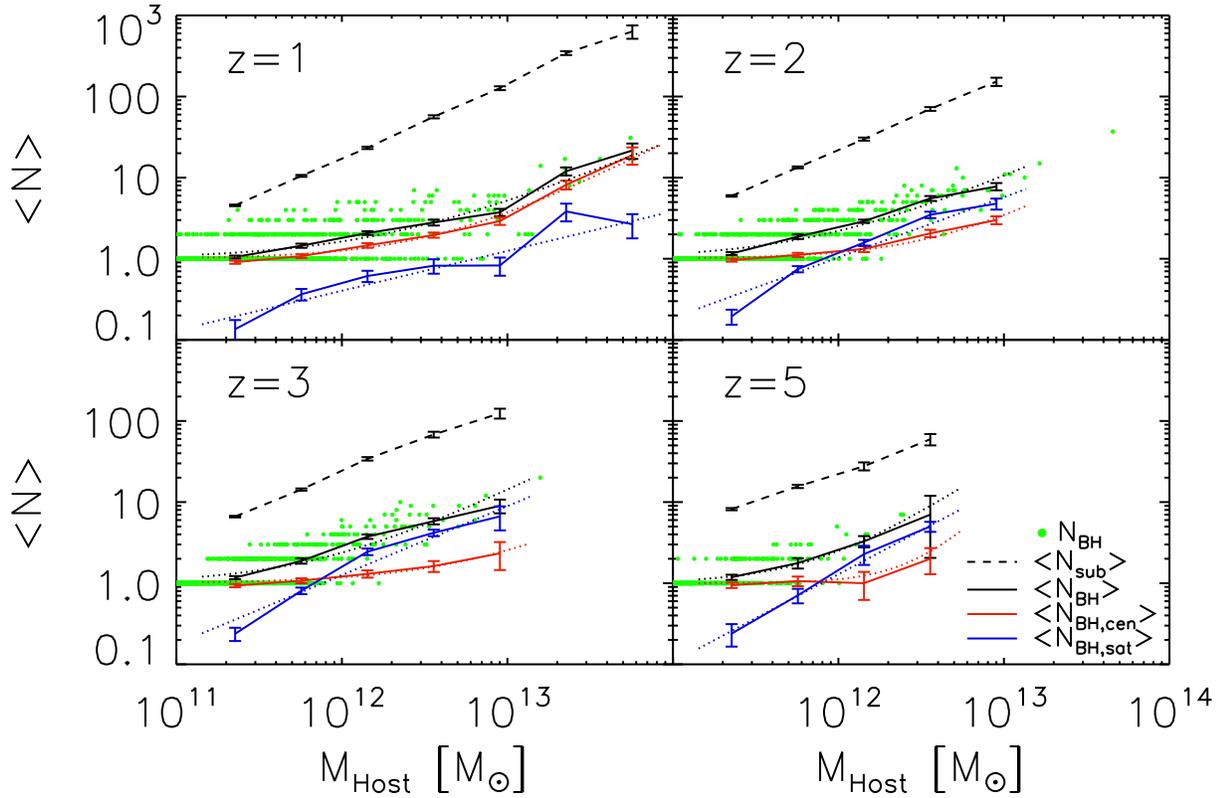}
\caption{Mean occupation number of subhalos (dashed black line), total black holes
  (solid black line), black holes in the central galaxy (red), and black holes
  in the satellite galaxies (blue).  We also show the number of BHs in each
  individual group (green dots), and the fits to the BH occupation number
  using Eqns. \ref{powerlaw}-\ref{powerlaw_sat} and Table \ref{HODparam} (dotted lines).}
\label{occupationnumber}
\end{figure*}
\begin{figure}
  \centering
  \includegraphics[width=8cm]{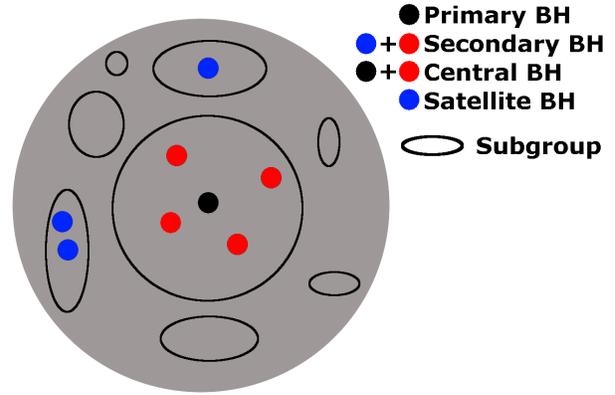}
  \caption{Cartoon representation of terminology used.  The grey circle
  represents a single large halo, with subhalos shown as black ellipses.  Primary BH: the most
  massive BH found in the halo (black).  Central BH: All BHs found in the
  central (most massive) subhalo (red+black).  Satellite BH: All BHs found in
  satellite subhalos (blue).  Secondary BH: Any non-primary BH (blue+red).}
  \label{definitions}
\end{figure}

The most basic component of the HOD model is the occupation number. In Figure
\ref{occupationnumber} we show the mean occupation number for both BHs (solid
black line) and subgroups (dashed black line) as a function of host halo mass,
as well as the the exact number of BHs found in each individual group (green
dots).  Note that these numbers are based on the full BH population with no
mass cut; see Section \ref{sec:cmf} for the BH mass distribution.  We also show the contributions to $\langle N_{\rm{BH}} \rangle$ arising
from BHs found in the central (i.e. most massive) galaxy (red line) and those
found in satellite galaxies (blue line).  We note that this is fundamentally
different from the traditional galaxy HOD model, in which `central' galaxy is
of course just one. Multiple black holes can be found in a central galaxy (at
least at low redshift as remnants of previous mergers) and therefore we do not
have this restriction. For clarity, a schematic representation of these
components (and some further subdivision we will discuss below) is shown in
Figure~\ref{definitions}.

In analogy with standard HOD models for the galaxy population we model the
total, central and satellite BH occupation number (where $ \langle N_{\rm{BH, tot}}
\rangle = \langle N_{\rm{BH,cen}} \rangle + \langle N_{\rm{BH,sat}} \rangle$) as:
\begin{equation}
  \langle N_{\rm{BH, tot}} \rangle = 1 + \left ( \frac{M_{\rm{Host}}}{M_{0}} \right )^{\alpha_{\rm{tot}}},
\label{powerlaw}
\end{equation}
\begin{equation}
  \langle N_{\rm{BH, cen}} \rangle = 1 + \left ( \frac{M_{\rm{Host}}}{M_{1}} \right)^{\alpha_{\rm{cen}}} ,
\label{powerlaw_cen}
\end{equation}
\begin{equation}
  \langle N_{\rm{BH, sat}} \rangle = \left ( \frac{M_{\rm{Host}}}{M_{2}} \right
  )^{\alpha_{\rm{sat}}} 
\label{powerlaw_sat}
\end{equation}
respectively. 
where $M_{\rm{Host}}$ is the halo mass of the host, $M_0, M_1$ and $M_2$ are
normalization constants which represent the host masses for which
we have a total of typically two black holes per host, two black holes in the
central galaxy and one in a satellite galaxy, respectively. Finally
  $\alpha_{\rm{tot}},\alpha_{\rm{cen}}$ and $\alpha_{\rm{sat}}$ are the exponents of the
  power law functions above. Note that Equations
  \ref{powerlaw}-\ref{powerlaw_sat} are not self consistent, but rather
  Equations \ref{powerlaw_cen}-\ref{powerlaw_sat} provide an alternative
  parameterization from Equation \ref{powerlaw}.

Standard galaxy HOD usually have a form $ \langle N_{\rm{gal, tot}}
\rangle = \langle N_{\rm{gal,cen}} \rangle + \langle N_{\rm{gal,sat}} \rangle$, here for
BHs $\langle N_{\rm{gal,cen}} \rangle$ has been replaced with a constant (equal to
one) to represent our seeding condition (which artificially imposes this
condition) and the power law forms $\langle N_{\rm{gal,sat}} \rangle$ is similar to
the ones above. Note again, here we have introduced an additional power law
for modelling the $N_{\rm BH, cen}$ which allow us to characterize the 
BH numbers in central galaxies.

Throughout this paper we refer to the most massive BH in a given group as the
`primary BH' while the remaining BHs are referred to as `secondary BHs'
(gained by merging with other BH-hosting halos; see Fig.\ref{definitions}).

Note this makes $\langle N_{\rm{BH,secondary}} \rangle = \left (
  \frac{M_{\rm{Host}}}{M_{0}} \right )^{\alpha_{\rm{tot}}}$ by definition (given
Equation \ref{powerlaw}).  This formulation is advantageous as
it can be used in clustering calculations in the same manner as the general
galaxy HOD \citep[see, e.g.][]{BerlindWeinberg2002, Kravtsov2004, Zheng2004}.

The function for $\langle N_{\rm{BH,cen}} \rangle$ also includes a constant (equal
to one) since BHs are seeded in the central subgroup.  The form of $\langle
N_{\rm{BH,sat}} \rangle$ lacks this constant since we do not seed subgroups with
BHs, and thus there need not be any BHs found in the satellite galaxies.  We
examine each $\langle N \rangle$ to its appropriate form
(Eqn. \ref{powerlaw}-\ref{powerlaw_sat}) based on halos at least twice the
threshold mass for seeding BHs (to avoid considering just-seeded halos).  The
results of these models are plotted on Figure \ref{occupationnumber} as dotted
lines.  We emphasize that these simple fits are intended to provide a
framework within which to get the typical number of BHs (total, satellite, and central) for the
mass ranges probed in our simulation, but care should be taken when
extrapolating to higher masses, particularly at high redshift where we have few
data points.

\begin{table}
\caption{Best fitting HOD parameters for Equations \ref{powerlaw}-\ref{powerlaw_sat}.}
\centering
\begin{tabular}{c c c c c}
\hline
\hline

 & \multicolumn{4}{c}{Redshift}\\

 & 1 & 2 & 3 & 5 \\

\hline

$\alpha_{\rm{tot}}$ & $ 0.82$ &$ 0.90$ &$ 0.98$ &$ 1.3$ \\
$M_{0}$ & $ 1.7 \times 10^{12}$ &$ 8.1\times
10^{11}$ &$ 7.2 \times 10^{11}$ &$ 7.3 \times 10^{11}$\\
\hline

$\alpha_{\rm{cen}}$ & $1.02$ & $ 1.1$ &$ 0.94$ &$ 2.0$\\
$M_{1}$ & $ 3.7 \times 10^{12}$ & $ 4.4 \times 10^{12}$ &$
6.2 \times 10^{12}$ &$ 2.9 \times 10^{12}$\\
\hline

$\alpha_{\rm{sat}}$ &$ 0.49$ &$ 0.74$ &$ 0.85$ &$ 1.1$\\
$M_{2}$ & $ 6.4 \times 10^{12}$ &$ 9.5 \times 10^{11}$ &$ 7.6 \times 10^{11}$ &$ 7.9 \times 10^{11}$ \\

\hline
\end{tabular}
\label{HODparam}
\end{table}

\begin{figure}
\centering
\includegraphics[width=8cm]{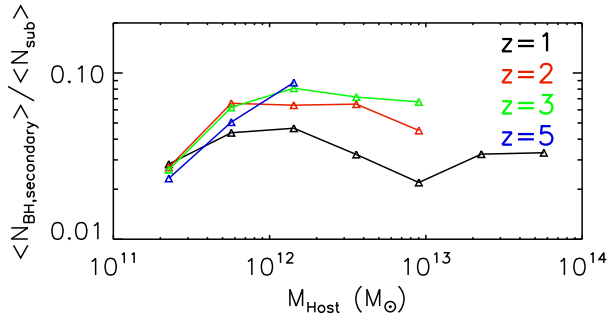}
\caption{Ratio of $\langle N_{\rm{BH,secondary}} \rangle$ to $\langle N_{\rm{subhalo}} \rangle$ at z=1 (black), 2
  (red), 3 (green), and 5 (blue).}
\label{ratioevolution}
\end{figure}

We see that $\langle N_{\rm{BH}} \rangle$ exhibits a general trend of mildly
decreasing slope ($\alpha$) with decreasing redshift, and that halos at low
redshift tend to have fewer BHs than halos with comparable mass at high
redshift. To further understand this effect, in Figure \ref{ratioevolution} we
show the ratio of $\langle N_{\rm{BH, secondary}} \rangle$ to $\langle
N_{\rm{subhalos}} \rangle$.  We note that $\langle N_{\rm{subhalos}} \rangle$
is sensitive to the mass threshold used to define a subhalo.  In this paper,
we consider any subgroup found by the SUBFIND algorithm (see section 2.3) to
be a subhalo.  However, we note that although the normalization of $\langle
N_{\rm{subhalos}} \rangle$ is sensitive to the mass threshold, the slope is
not.  To avoid delving into the issue of subgroup definitions and the
model-dependencies therein, we limit ourselves to investigating how the ratio
evolves, which does not exhibit significant dependence on the mass threshold.
In low mass halos, the ratio does not evolve with redshift, and at each
redshift there are fewer BHs per subhalo for the low-mass halos. This is expected since these
are halos close to the threshold mass for seeding a BH, and few will have
undergone a merger that can lead to a secondary BH.  However, we note that
more massive halos at low redshift tend to have fewer BHs per subhalo (and
fewer BHs in general) than the corresponding halos at high redshift.  This
decrease is likely a result of changes in the halo and BH merger rates.
Because the number of BHs in a halo depends both on the rate at which new BHs
enter the halo (via halo mergers) and the rate at which BHs within the halo
merge with each other, a decrease in the halo merger rate relative to the BH merger rate
would explain the decreased number of BHs.

\begin{figure*}
\centering
\includegraphics[width=14cm]{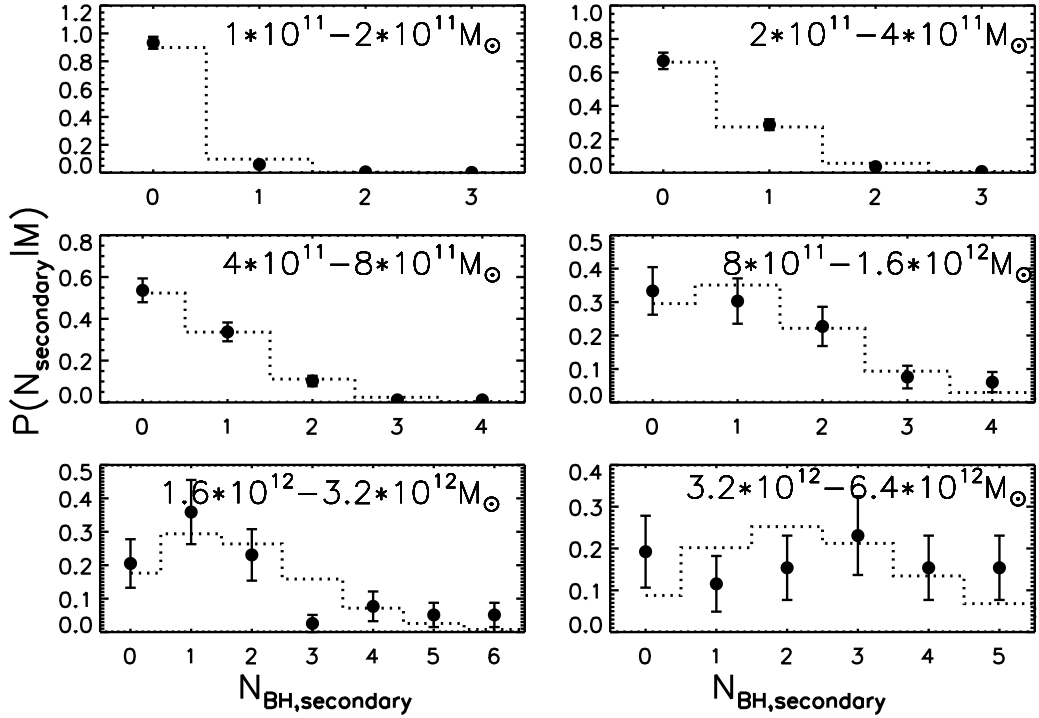}
\caption{The probability distribution of $N_{\rm{BH,secondary}}$ for several host
  halo mass ranges.  Filled circles
  show the results of our simulation (with Poisson error bars), and the dotted
  line shows a Poisson distribution centered about $\langle N_{\rm{BH,secondary}} \rangle$ for
  the given host halos.  }
\label{probdist}
\end{figure*}

To characterize the scatter in $N_{\rm{BH}}$, in Figure \ref{probdist} we plot
the probability distribution $P(N_{\rm{BH,secondary}}|M)$ (the probability of
a halo of mass M hosting $N_{\rm{BH,secondary}}$ BHs in addition to the
initially-seeded, primary BH) for several ranges of host halo masses (filled
circles with Poisson error bars).  For comparison, we show a Poisson
distribution with the same mean as the specified host mass range (dotted
line).  We see that the distribution of $N_{\rm{BH,sat}}$ is extremely close
to the Poisson distribution for the lower mass ranges (where our simulation
has a large sample of halos).  Even for high mass halos, where our
statistics are poor, it appears largely consistent with a Poisson distribution.
We note that we have chosen to plot the probability distribution for the
secondary BHs rather than the total number of BHs since the existence of the
primary BH is a condition enforced by our simulation which distorts $P(N|M)$
away from a Poisson distribution (by removing the possibility of $N_{\rm{BH}}
= 0$).  However, this is an expected effect from the model for seeding BHs
within our simulation, and we emphasize that the the physically-significant
$N_{\rm{BH,secondary}}$ is well fit by a Poisson distribution about $\langle
N_{\rm{BH,secondary}} \rangle$.  We also note that although we have only
plotted the distribution of $N_{\rm{BH,secondary}}$, we have also found that
$N_{\rm{BH,cen}}-1$ (to avoid inclusion of the model-imposed primary BH) and $N_{\rm{BH,sat}}$
both follow an approximate Poisson distribution as well.

\subsection{Black Hole Conditional Mass Function}
\label{sec:cmf} 

\begin{figure*}
\centering
\includegraphics[width=15cm]{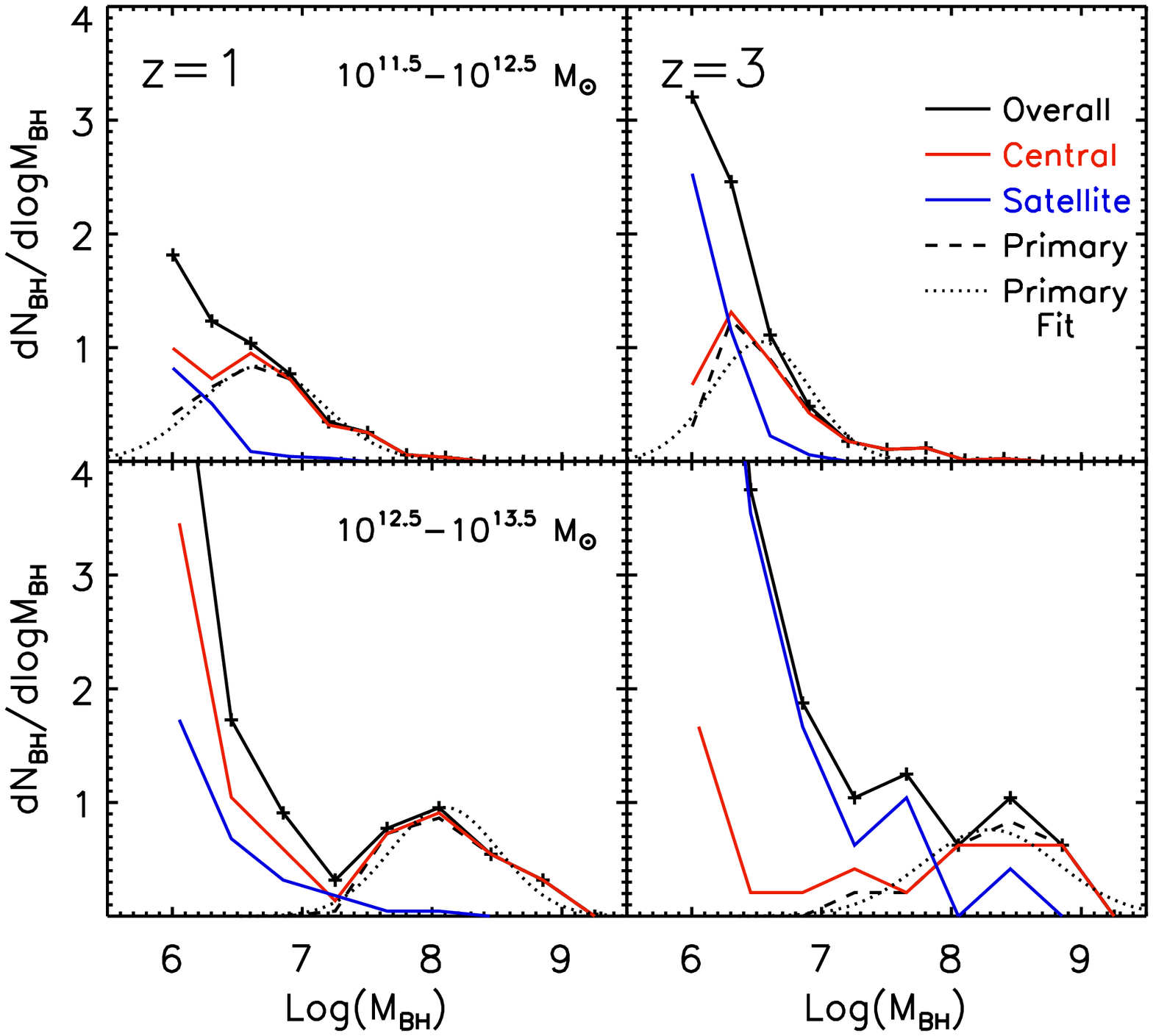}
\caption{Distribution of BH masses in host halos of three different mass
  ranges at z=1 and 3.  Note: Only a single group above $10^{13} M_\odot$
  exists at z=3, so that distribution has not been plotted.}
\label{CMF}
\end{figure*}

Another important facet of the model is the mass distribution of BHs
populating halos.  To investigate this we produce a conditional mass function
(CMF), similar to the conditional luminosity function done for galaxies
\citep[see, e.g.][]{Yang2003}.  For this, we use $\frac{dN_{\rm{BH}}}{d\log M_{\rm{BH}}}$,
the mean number of BHs per logarithmic BH mass bin found in halos of mass
$M_{\rm{Host}}$.  We plot this quantity in Figure \ref{CMF} for several host halo
mass ranges at z=1,3.  Here we see that in low mass halos, regardless of
redshift, the BHs tend to be close to the seed mass, as expected.  In higher
mass halos, we find the distribution to be bimodal.  The primary BH mass
is distributed about a $M_{\rm{Host}}$-dependent peak, while the remaining
BHs follow an approximate power law, with the majority of secondary BHs being
near the seed mass.  The secondary BH CMF is also mass-dependent, with the CMF
in lower mass halos dropping faster with increasing $M_{\rm{BH}}$ than in
lower mass halos.  The secondary mass distribution always peaks at the seed
mass, however, regardless of redshift or host halo mass.

The conditional mass function for the primary (i.e. most-massive) black hole
can be approximated by a Gaussian distribution (in $\log_{10}(M_{\rm{BH}})$) with an integrated area of one
\begin{equation}
\frac{dN_{\rm{BH,primary}}}{d \log M_{\rm{BH}}} = \frac{1}{\sqrt{2\pi \sigma^2}}
e^{-\frac{(\log_{10}(M_{\rm{BH}})-\mu)^2}{2\sigma^2}}.
\label{eqn:primary}
\end{equation}
The best fitting parameters for this function are provided in Table
\ref{CMFparam}.  Note there is only a single group above $10^{13} M_\odot$ at
z=3, so no fitting parameters are given for that mass range.  From these fits
we can see that the typical primary BH mass grows roughly proportionally to
the host mass range, and is approximately independent of redshift.
Furthermore, the CMF for secondary BHs is reasonably well-fit by a simple
power law
\begin{equation}
\frac{dN_{\rm{BH,secondary}}}{d\log M_{\rm{BH}}} = \left( \frac{M_{\rm{BH}}}{M_{0\rm{,Host}}}
\right)^\alpha,
\label{eqn:secondary}
\end{equation}
the parameters for which are given in Table \ref{CMFparam} for both
z=1 and z=3.  Overall, the clear trend is for the secondary BHs in small halos
to be more strongly concentrated near the seed mass, while larger halos are
more likely to have more massive secondary BHs.  We emphasize that these
trends are derived from the distributions for the mass ranges probed within
the simulation, but are not reliable if used for masses at or below the seed
mass. We also note that $\mu$ is roughly proportional to the host mass, so the
use of finite binsizes in $\log_{10}\left (M_{\rm{host}} \right )$ increases
$\sigma$ above the ideal values for fixed halo masses.  

\begin{figure*}
\centering
\subfigure{
\includegraphics[width=8cm]{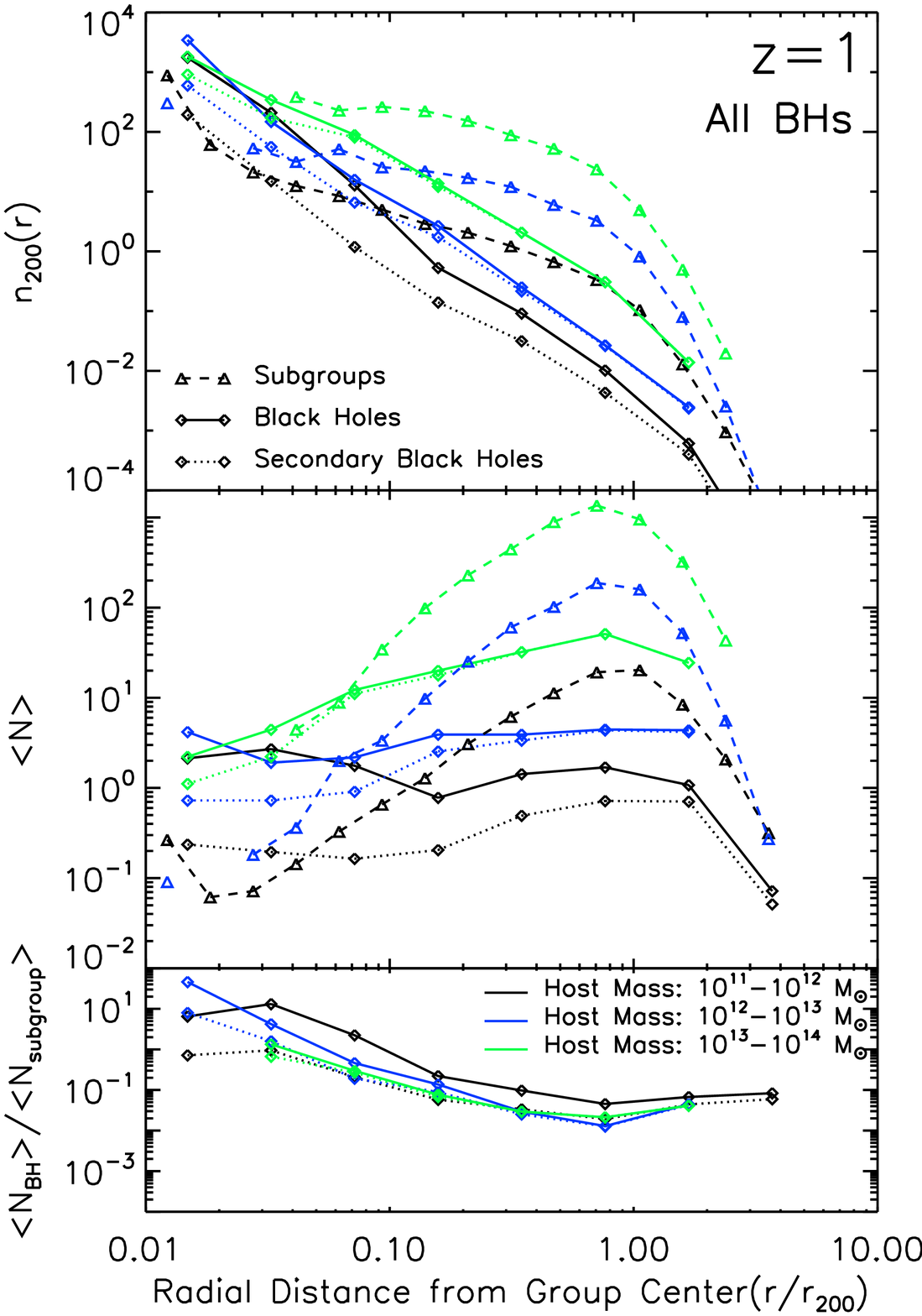}
}
\subfigure{
\includegraphics[width=8cm]{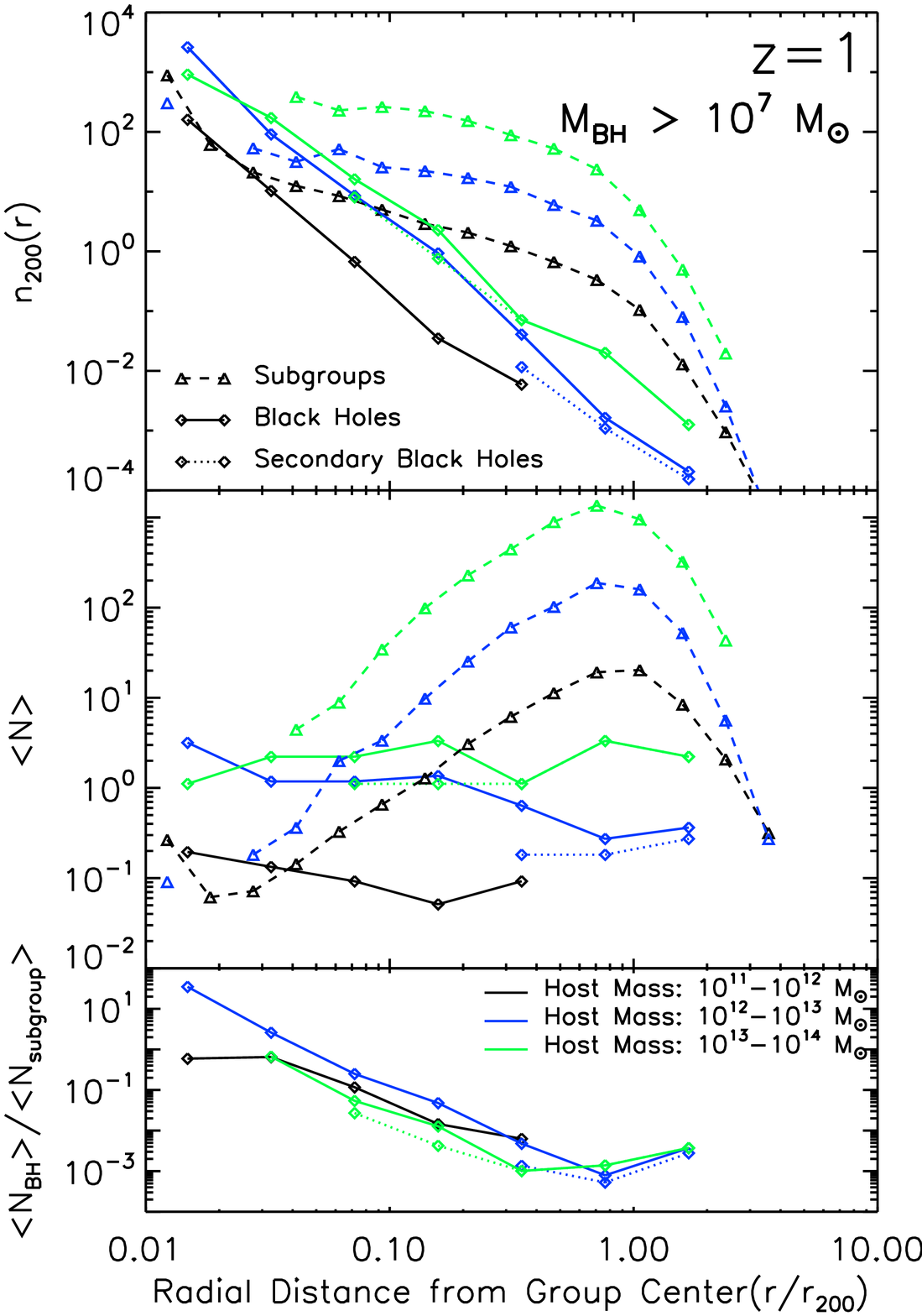}
}
\caption{\textit{Top:} Number density of subhalos (dashed
  lines), BHs (solid lines) and secondary BHs (dotted lines) in units of
  $R_{200}^{-3}$ for all BHs (left) and for BHs above $10^7 M_\odot$ (right).  \textit{Middle:} Mean number of BHs (solid lines) and subhalos (dashed
  lines) per radial bin at z=1. \textit{Bottom:} Ratio of BHs to subhalos as a function of radial distance
  from group center.   In all plots, color represents the host group mass
  range (black - $10^{11}-10^{12} M_\odot$;
  blue - $10^{12}-10^{13} M_\odot$; green - $10^{13}-10^{14}
  M_\odot$).}
\label{radialdist}
\end{figure*}

\begin{table}
\caption{Fit for conditional mass function (Equations \ref{eqn:primary} and \ref{eqn:secondary})}
\begin{tabular}{c c c c c c c}

\hline
\hline

 & & \multicolumn{4}{c}{\underline{Host Mass Range $\left ( \log_{10} \left (\frac{M}{M_\odot}
 \right ) \right )$}} \\
 & & $11.5-12.5$ & $12-13$ & $12.5-13.5$ & $13-14$ \\

\hline
\multicolumn{5}{l}{\underline{Primary BHs:}} \\
\underline{z=1} 
    & $\mu$ &  $6.68$ & $7.48$ & $8.10$ & $8.77$ \\
    & $\sigma$ & $.48$ & $.50$ & $.42$ & $.31$ \\
\underline{z=3}
    & $\mu$ & $6.54$ & $7.28$ & $8.30$ & N/A\\
    & $\sigma$ & $.38$ & $.75$ & $.52$ & N/A\\
\hline

\multicolumn{5}{l}{\underline{Secondary BHs:}} \\
\underline{z=1} & $\alpha$ & $-1.51$ & $-1.09$ & $-1.01$ & $-1.0$ \\ 
    \multicolumn{2}{r}{$\log_{10}{\left (M_{0\rm{,Host}} \right)}$} & $6.12$&
    $6.42$& $6.75$& $7.21$\\
\underline{z=3} & $\alpha$ & $-1.79$ & $-1.15$ & $-.71$ & N/A\\
    \multicolumn{2}{r}{$\log_{10}{\left (M_{0\rm{,Host}} \right)}$} & $6.29$&
    $6.74$& $7.37$&  N/A  \\

\hline
\end{tabular}
\label{CMFparam}
\end{table}

\subsection{Black Hole Radial Distribution}
\label{sec:radialdist}
Another aspect of the HOD model is the spatial distribution of black holes
within their parent halos.  Although we have already analyzed how they
populate subhalos within their hosts, it may also be useful to understand how
BHs populate halos for which the subhalos have not been identified.  For this,
in Figure \ref{radialdist} we show the radial distribution of subhalos (dashed
lines), BHs (solid lines), and secondary BHs (dotted lines) at
redshift 1 for host groups separated into three mass bins (black:
$10^{11}-10^{12} M_\odot$, blue: $10^{12}-10^{13} M_\odot$, green:
$10^{13}-10^{14} M_\odot$), expressed as both a number density in units of
$R_{200}^{-3}$ (top), and simply as a number per radial bin (middle).  In the
lower plot we show the ratio of BHs to subhalos as a function of radial
distance from the center.  We do
this using the complete BH population (left) and using only BHs above $10^7
M_\odot$ (right).  Although only shown for $z=1$, we find very similar
results for $z=3,5$ as well.  Thus, although the number of BHs found in halos
of a given mass changes with redshift (as seen in Table \ref{HODparam}), the
manner in which they are distributed within the halos remains approximately
the same.

Figure \ref{radialdist} shows that for any host mass range, black holes are
substantially more centrally-concentrated than the subhalos, as predicted by
the black hole clustering properties in \citet{DeGraf2010Clustering}.  In
fact, they follow a fundamentally different profile which can be modeled by a
simple power law rather than the more typical NFW profile \citep[again in
keeping with the results of][]{DeGraf2010Clustering}.  This increased
concentration is a result of mergers between BH-hosting subgroups, typically
between the central subgroup and a satellite.  Because the central subgroup
absorbs the satellite subgroup, the concentration of the subgroups does not
increase, but that of the BHs will, since the BH will survive for a
non-negligible time before a merging by the primary BH (and if it does, this
is set by dynamical friction which is solved for in the simulations).
Essentially, it is the existence of non-primary central BHs (see Figure
\ref{definitions}) which produces this increased central concentration.  For
further details see also \citet{DeGraf2010Clustering}.

To provide a means for modeling the spatial distribution of black holes within
their parent halos (when populating halos directly rather than populating the
central and satellite galaxies), we provide a simple fit to the radial profile
(for $r < 2 \times r_{\rm{200}}$) in form of a  power law
\begin{equation}
n_{200} (r) \propto \left (\frac{r}{r_{200}} \right )^\beta ,
\end{equation}
which should be normalized to the occupation number found with Equation
\ref{powerlaw}.  The values for
$\beta$ are listed in Table \ref{RadialParameters}.  We note that for low-mass
halos (below $\sim 10^{12} M_\odot$), the BHs are distributed roughly
uniformly (in logarithmic bins) out to $\sim R_{200}$, while the most massive
halos tend to have more BHs further from the center (though the density of BHs
nonetheless decreases rapidly with radial
distance).   
We also note that when approaching the group center, the ratio of BHs to
subhalos tends to grow faster for low-mass halos than for more massive ones.
Although this means that BHs in low mass halos are typically more centrally
concentrated, we note that this is a result of the increased importance of the
primary BH due to the smaller BH occupation number.  However, we note that the
ratio of secondary BHs to subhalos (dotted lines) is constant, regardless of
host halo mass.

We also plot the radial distribution for massive BHs ($M_{\rm{BH}} > 10^7
M_\odot$, right column) in Figure \ref{radialdist}.  Here we see that in
addition to being less common, the more massive BHs are more centrally
concentrated than the less massive BHs.  This is expected since the massive
BHs are almost exclusively primary BH, and only rarely are they satelite
BHs (see Figure \ref{CMF}), and thus they should be more highly concentrated
toward the center of the group.

\begin{table}
\caption{Radial distribution parameter for functional form $n_{200}(r)
  \: \propto \left (\frac{r}{r_{200}} \right )^\beta$}
\centering
\begin{tabular}{c c c}
\hline
\hline

Host Mass & $\beta$ \\

\hline

$10^{11}-10^{12} M_\odot$ & -3.14 \\
$10^{12}-10^{13} M_\odot$ & -2.89 \\
$10^{13}-10^{14} M_\odot$ & -2.20 \\

\hline
\end{tabular}
\label{RadialParameters}
\end{table}

\begin{figure}
\centering
\includegraphics[width=8cm]{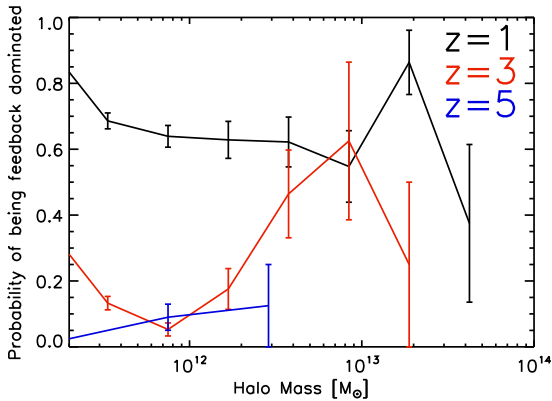}
\caption{The fraction of groups which are feedback suppressed defined as
  $M_{\rm{BH}}$ above the quoted scatter of the $M-\sigma$ relation of \citet{DiMatteo2008}
  for z=1,3,5.  }
\label{feedbacksuppression}
\end{figure}

\subsection{Black Hole Feedback Suppression}
\label{sec:feedback}
One final aspect of our analysis we would like to characterize is the mass of
dark matter halos for which black hole feedback has been significant (and,
for example, has been responsible for shutting down star formation in its halo.)
In our model, and in others in the literature it has been shown that feedback
from a central BH can create outflows that are able to expel a substantial
amount of the gas from the host galaxy, thereby suppressing further growth of
the BH, reproducing the $M-\sigma$ relation, and shutting down star
formation \citep{2005Natur...433..604D,2005MNRAS.361..776S,2006ApJS..163....1H,2007ApJ...669...67H,DiMatteo2008}

Here we fit our simulations to provide a probabilty for a given mass halo
to have been significantly affected by BH feedback (i.e. feedback dominated).
\citet{DiMatteo2008} showed that our model reprduced the observed 
$M-\sigma$ relation and it does so as a result of BH feedback.
We use the $M-\sigma$ relation from our simulation to obtain the black hole
mass and the respective halo mass for which BHs are feedback dominated.

In Figure \ref{feedbacksuppression} we show the fraction of groups in each
halo mass bin whose BH is large enough to be considered `feedback suppressed'
(where the feedback is strong enough to suppress further growth),
using the condition that any $M_{\rm{BH}}$ above the quoted scatter of the
$M-\sigma$ relation of \citet{DiMatteo2008} is feedback suppressed.
We note that the exact choice of cutoff threshold has a mild effect on the
overall amplitude (i.e. the exact fraction of feedback suppressed halos), but
the general trends are insensitive to the cutoff criteria. We find that at
high redshift, very few halos have sufficiently large BHs to be feedback
regulated, with larger halos being slightly more likely to have reached it
than smaller halos.  As time passes, the BHs in the high mass halos become
more likely to become feedback regulated, and gradually the less-massive
halos begin to become suppressed as well.

\section{Conclusions}
\label{sec:conclusions}

\begin{itemize}

\item The BH occupation number is well-described by the functional form
  $\langle N_{\rm{BH}} \rangle = 1 + \left ( \frac{M}{M_{0}} \right
  )^{\alpha_{\rm{tot}}}$ for directly populating dark matter halos.  Alternatively,
  separate occupation numbers can be obtained for BHs in central and
  satellite galaxies (Eqns. \ref{powerlaw_cen}-\ref{powerlaw_sat}) to populate
  subgroups in an N-body simulation (or galaxy HOD model).  

\item In general, $\langle N_{\rm{BH}} \rangle$ typically follows
  $\langle N_{\rm{subhalo}} \rangle$ fairly consistently, suggesting BHs populate subhalos similarly
  regardless of host halo mass.  At low redshift, however, we find there are
  fewer BHs in the hosts (both total and relative to $N_{\rm{subhalo}}$), particularly in moderate-mass halos, presumably as a
  result of the changing merger rates of both halos and the BHs within halos.  

\item The scatter in $\langle N_{\rm{BH}} \rangle$ is well described by a
  single primary BH
  and a number of secondary BHs that follow a Poisson distribution about the
  mean secondary occupation number 
  $\langle N_{\rm{BH,secondary}} \rangle = \left ( \frac{M_{\rm{Host}}}{M_{0}} \right
  )^{\alpha_{\rm{tot}}}$.  We also find that the central and satellite occupation
  numbers follow approximate Poisson distributions.

\item The conditional mass function for the primary BH peaks around a BH mass
  strongly correlated with $M_{\rm{Host}}$.  The secondary BH mass
  distribution is peaked at the seed mass, and falls off as a power law in
  $M_{\rm{BH}}$.  The power law is steepest for smaller host halos, such that more
  massive halos have a wider spread of BH masses, as expected.

\item The spatial distribution of black holes within halos is fundamentally
  different from that of subhalos, tending to follow a power law rather than
  an NFW profile, leading to a significantly stronger central concentration of
  BHs relative to both subhalos and the underlying dark matter distribution.
  This increased concentration supports the predictions made in
  \citet{DeGraf2010Clustering}, though more direct investigation into our
  HOD-predicted correlation function will be investigated in an upcoming paper.

\item For a given host halo mass, the spatial distribution of black holes
  does not evolve with redshift.  Thus although the number of
  BHs per host halo changes with $z$, how they are distributed within these
  halos remains generally the same.

\item At high redshift, few BHs are sufficiently massive to reach the observed
  $M-\sigma$ relation.  When moving to lower redshifts, the more massive halos
  are generally the first to reach the $M-\sigma$ relation, with the
  lower-mass halos reaching the relation last.  This suggests that the larger
  halos become suppressed by BH feedback at early time, and only at late
  times do the smaller halos begin to experience these suppressing effects of
  BH feedback.

\item We have provided best fit parameters for the mean occupation number
  $\langle N_{\rm{BH}} \rangle$ as a function of host group mass, as well as
  the BH-mass and spatial distribution functions within these halos to provide
  the necessary information to populate dark matter halos with black holes.
  Alternatively, we have provided the mean occupation number of BHs found in
  the central and satellite galaxies ($\langle N_{\rm{BH,cen}} \rangle$ and
  $\langle N_{\rm{BH,sat}} \rangle$, respectively) to provide the necessary
  information for directly populating subgroups with BHs.

\end{itemize}

\section*{Acknowledgments}

We would like to thank Michael Busha for his suggestion to investigate when
halos become black hole feedback dominated.  This work was supported by the
National Science Foundation, NSF Petapps, OCI-0749212 and NSF AST-0607819.
The simulations were carried out at the NSF Teragrid Pittsburgh Supercomputing
Center (PSC). D.N. was supported in part by the NSF grant AST-1009811, by 
NASA ATP grant NNX11AE07G, and by Yale University.  Z.Z. gratefully acknowledges support from Yale Center for Astronomy and Astrophysics through a YCAA fellowship.

 \bibliographystyle{mn2e}	
 \bibliography{astrobibl}	

\end{document}